# Correlated Electronic Structure and Density-Wave Gap in Trilayer Nickelate La$_4$Ni$_3$O$_{10}$

X. Du[1], Y. D. Li[1], Y. T. Cao[2,3], C. Y. Pei[4], M. X. Zhang[4], W. X. Zhao[1], K. Y. Zhai[1], R. Z. Xu[1], Z. K. Liu[4,5], Z. W. Li[2], J. K. Zhao[3], G. Li[4], Y. L. Chen[4,5,6,†], Y. P. Qi[4,5,7,†], H. J. Guo[3,†], and L. X. Yang[1,8,†]

[1]*State Key Laboratory of Low Dimensional Quantum Physics, Department of Physics, Tsinghua University, Beijing 100084, China*
[2]*School of Physical Science and Technology, Lanzhou University, Lanzhou 730000, China*
[3]*Songshan Lake Materials Laboratory, Dongguan, Guangdong 523808, China*
[4]*School of Physical Science and Technology, ShanghaiTech University, Shanghai 201210, China*
[5]*ShanghaiTech Laboratory for Topological Physics, Shanghai 200031, China*
[6]*Department of Physics, Clarendon Laboratory, University of Oxford, Parks Road, Oxford OX1 3PU, UK*
[7]*Shanghai Key Laboratory of High-resolution Electron Microscopy, ShanghaiTech University, Shanghai 201210, China*
[8]*Frontier Science Center for Quantum Information, Beijing 100084, China*

*Emails: LXY: lxyang@tsinghua.edu.cn; YLC: yulin.chen@physics.ox.ac.uk; YPQ: qiyp@shanghaitech.edu.cn; HJG: hjguo@sslab.org.cn.*

**The discovery of pressurized superconductivity at 80 K in La$_3$Ni$_2$O$_7$ officially brings nickelates into the family of high-temperature superconductors, which gives rise to not only new insights but also mysteries in the strongly correlated superconductivity. More recently, the sibling compound La$_4$Ni$_3$O$_{10}$ was also shown to be superconducting below about 25 K under pressure, further boosting the popularity of nickelates in the Ruddlesden-Popper phase. In this study, combining high-resolution angle-resolved photoemission spectroscopy and *ab initio* calculation, we systematically investigate the electronic structures of La$_4$Ni$_3$O$_{10}$ at ambient pressure. We reveal a high resemblance of La$_4$Ni$_3$O$_{10}$ with La$_3$Ni$_2$O$_7$ in the orbital-dependent fermiology and electronic structure, suggesting a similar electronic correlation between the two compounds. The temperature-dependent measurements imply an orbital-dependent energy gap related to the density-wave transition in La$_4$Ni$_3$O$_{10}$. By comparing the theoretical pressure-dependent electronic structure, clues about the superconducting high-pressure phase can be deduced from the ambient measurements, providing crucial information for deciphering the unconventional superconductivity in nickelates.**

Since its discovery, high-temperature superconductivity (HTSC) in cuprates remains one of the most challenging mysteries in condensed matter physics [1,2]. To advance the understanding of this problem, it is highly pursued to search for analogs to cuprates with similar superconducting and normal-state properties. Nickelates are believed to be promising candidates for high-temperature superconductors with strong electronic correlation. Unconventional superconductivity has been discovered in infinite-layer [3-5] and quintuple-layer [6] nickelates, in which $3d$ electrons of Ni dominate the electronic states near the Fermi level ($E_F$). In particular, $Ni^+$ exhibits an effective electronic configuration of $3d^9$, strikingly resembling $Cu^{2+}$ in cuprates [7,8]. Moreover, the electronic correlation likewise plays a pivotal role in the electronic structure of nickelate superconductors [9]. Such similitude makes nickelates an attractive platform for exploring HTSC. However, the synthesis of these nickelates is challenging and the superconductivity is realized only in thin films with $T_c$ far below the criterion of HTSC regardless of the application of pressure [3,5].

Recently, HTSC with $T_c$ up to 80 K was discovered in $La_3Ni_2O_7$ bulk compound under pressures between 14.0 GPa and 43.5 GPa [10-13]. Likewise, the signature of superconductivity under pressure with the maximum $T_c$ between 20 K and 30 K has been observed in the sister compound $La_4Ni_3O_{10}$ [14-16]. With the chemical formula of $La_{n+1}Ni_nO_{3n+1}$, both compounds belong to the Ruddlesden-Popper (RP) phase, which can be viewed as the parent phase of the infinite-layer and quintuple-layer nickelates. They exhibit extraordinary properties at ambient pressure, including orbital-dependent electronic correlation, non-Fermi liquid behavior, and the interplay with density-wave states [12,17-19]. Many theoretical understandings of the electronic structure and superconductivity of the bilayer and trilayer nickelates have been proposed with the focus on the correlated electronic structure of the $d_{x^2-y^2}$ and $d_{z^2}$ orbitals [20-33]. While it is challenging to experimentally investigate the electronic structure at high pressure, the experiments at ambient pressure will provide crucial information. Previous angle-resolved photoemission spectroscopy (ARPES) measurements have revealed the basic electronic structure and electronic correlation in $La_3Ni_2O_7$ [17,34] and $La_4Ni_3O_{10}$ [35], with more details awaiting further experimental exploration. Furthermore, a comparing investigation of the electronic structures of the bilayer and trilayer nickelates will offer important insights into their normal-state and superconducting properties.

In this work, we systematically investigate the electronic structure of $La_4Ni_3O_{10}$ at ambient pressure with high-resolution ARPES in comparison to La$_3$Ni$_2$O$_7$ and explore the effect of pressure with *ab initio* calculation. The experimental and calculated band structure of La$_4$Ni$_3$O$_{10}$ is very similar to that of La$_3$Ni$_2$O$_7$. The low-energy electronic structure is dominated by Ni $3d_{x^2-y^2}$ and $3d_{z^2}$ orbitals. In particular, the $3d_{z^2}$ orbitals form flat bands near $E_F$, which contribute significantly to the density of states (DOS). At high pressure, our *ab initio* calculation suggests that the bandwidth increases by about 20%, while the band dispersions near $E_F$ show a minor change. Furthermore, in the energy bands with $d_{x^2-y^2}$ and $d_{z^2}$ hybridization, we observe a gap showing Bardeen-Cooper-Shrieffer (BCS) type temperature-dependence below the charge/spin-density wave (CDW/SDW) transition temperature of La$_4$Ni$_3$O$_{10}$. More interestingly, the gap amplitude shows an anisotropy that is correlated with the strength of the hybridization between the $d_{x^2-y^2}$ and $d_{z^2}$ orbitals. Our comparing study of the bilayer and trilayer nickelates will help construct a unified understanding of the normal-state and superconducting picture of the unconventional HTSC in nickelates of the RP phase.

The crystal structures of La$_4$Ni$_3$O$_{10}$ and La$_3$Ni$_2$O$_7$ are shown in Figs. 1(a) and 1(b), respectively. As sister compounds in the RP phase, both materials feature layered structures made of corner-sharing Ni-O octahedron layers [10,36]. In contrast to the equivalent Ni-O layers in La$_3$Ni$_2$O$_7$, La$_4$Ni$_3$O$_{10}$ harbors two sets of inequivalent Ni-O layers. At ambient pressure, the vertical axes of the octahedrons cant by an angle of about 8° (6°) with respect to the *c* axis in La$_4$Ni$_3$O$_{10}$ (La$_3$Ni$_2$O$_7$). A pressure above about 13 GPa aligns the axes and changes the crystal symmetry, which is believed to be crucial for the superconductivity [36]. Laue measurement [Fig. 1(c)] of the (001) surface confirms the high quality of the La$_4$Ni$_3$O$_{10}$ crystals used in our work. The magnetic susceptibility decreases with decreasing temperature followed by a sudden drop (enhancement) of the in-plane (out-of-plane) component near 132 K [Fig. 1(c)]. Correspondingly, the specific heat shows a cusp near 132 K [Fig. 1(d)], in nice agreement with previous reports [37,38], which has been ascribed to an intertwined density-wave transition with both charge and magnetic characters [39].

Figures 2 (a-d) present the measured band structure of La$_4$Ni$_3$O$_{10}$. As shown in Fig. 2(a), the Fermi surface (FS) is mainly made up of three sets of bands. The α and β bands compose a nearly-circular sheet and a larger rounded-square sheet around the $\bar{\Gamma}$ point, respectively, which are better visualized in the laser-ARPES data in Fig. 2(b). Along the $\bar{\Gamma}\bar{X}$ direction [Fig. 2(c)], the α and β bands are nearly degenerate, while the γ band approaches $E_F$ with a flat band top. The flat γ band is better resolved along the $\overline{\Gamma'S}$ direction in the second Brillouin zone [the red dashed line in Fig. 2(a), also see Supplemental Materials [40]]. The overall electronic structure is in good agreement with the previous ARPES experiment [14,35].

In Figs. 2(e-h) we compare the band structures of $La_4Ni_3O_{10}$ and $La_3Ni_2O_7$. Despite their distinct number of Ni-O octahedron layers, both their Fermiology and energy dispersions are analogous (more data are presented in Supplemental Materials [40]). In particular, both FSs measured at 98 eV are consisted of parallel FS sheets with strong spectral weight, providing an electronic basis for the FS nesting. The observed similarity between the electronic structures of $La_4Ni_3O_{10}$ and $La_3Ni_2O_7$ is, however, out of the expectation since the inequivalent Ni-O layers in $La_4Ni_3O_{10}$ are supposed to contribute additional bands. The absence of additional bands compared to $La_3Ni_2O_7$ might be due to the matrix element effect or a delicate band-splitting issue [35]. Indeed, the experimental electronic structure of a $La_5Ni_4O_{13}$ thin film, another member of the RP phase nickelates, also resembles those of the two materials in our study [41,42].

To further comprehend the electronic structure of $La_4Ni_3O_{10}$, we perform detailed band structure calculations in Fig. 3 (and a comparison with $La_3Ni_2O_7$ in Supplemental Materials [40]). Figure 3(a) shows the *ab initio* calculation of the band structure projected onto atomic orbitals (Supplemental Materials [40]). The bands near $E_F$ are mainly composed of Ni 3$d$ orbitals hybridized with O 2$p$ orbitals. Due to the crystal-field splitting, the $e_g$ bands locate near $E_F$ and separate from the $t_{2g}$ bands in energy. The trilayer structure induces three sets of energy bands, as can be better visualized near the Γ and Z points. Due to the inter-layer coupling, the $d_{z^2}$-orbital derived bands split into bonding, non-bonding, and anti-bonding bands with distinct energies, while the in-plane $d_{x^2-y^2}$-orbital derived bands show a much smaller splitting.

Based on these characteristics, we carry out Wannier orbital calculation, which yields two sets of Wannier orbitals involving Ni $3d_{z^2}$ – O $2p_z$ and Ni $3d_{x^2-y^2}$ – O $2p_{x/y}$ hybridized atomic orbitals, respectively [Fig. 3(b)]. By comparing with the experimental band structure, we determine the renormalization factors of the $d_{z^2}$ and $d_{x^2-y^2}$ orbitals to be 5 and 3, respectively (Supplemental Materials [40]), which reflects the more-localized nature of the $d_{z^2}$ orbitals [27]. We also calculate the FS with momentum-dependent weight of atomic orbitals as shown in Fig. 3(c). The calculated and experimental results agree well if the $d_{z^2}/d_{x^2-y^2}$-derived bonding bands and the $d_{x^2-y^2}$-derived anti-bonding bands are considered [35] (Supplemental Materials [40]). It is noteworthy that the α and β bands are mixtures of the two $e_g$ orbitals ($d_{z^2}$ and $d_{x^2-y^2}$), with a pure $d_{x^2-y^2}$ character only along the $\Gamma X$ direction. By contrast, the γ band consists exclusively of $d_{z^2}$ character.

The electronic structure is also calculated for the pressurized phase at 30.5 GPa [Figs. 3(d-f)] [36]. As shown in Fig. 3(d), the overall band structure of the high-pressure phase coincides with that of the ambient phase. We note that the bandwidth increases by about 20%, while the band dispersions near $E_F$ barely change with increasing pressure, except that the overlap and hybridization between the bonding and non-bonding $d_{z^2}$ bands reduce near $\Gamma$ but increase near $Z$. Due to the alignment of the Ni-O octahedrons under pressure, the orbital components of the Wannier orbitals appear to be unmixed, forming inter- and intra-layer σ bonds, which are analogous to the situation in $La_3Ni_2O_7$ and favor the pressurized superconductivity [10]. In the FS, both the nearly-circular α-sheet and the large round-square-like β-sheet show minor change, while the small hole pocket from the hybridization between the bonding and non-bonding $d_{z^2}$ bands evolves into two small anisotropic pockets at Γ. It is worth noting that the small overlap between the bonding and non-bonding $d_{z^2}$ bands in the calculation may lead to the mutable topology of FS with pressure and electron doping.

Figure 4 presents the temperature-dependent laser-ARPES spectra along the $\bar{\Gamma}\bar{S}$ and $\bar{\Gamma}\bar{X}$ directions [dashed arrows in Fig. 2(b)], respectively. As shown in Fig. 4(a), the spectral weight of the α band reduces as approaching $E_F$ without a quasiparticle peak at low temperatures along the $\bar{\Gamma}\bar{S}$ direction. With increasing temperature, the spectral weight near $E_F$ increases [Figs. 4(b-d)] and the leading edge of the energy distribution curves (EDCs) gradually shifts towards $E_F$ [Figs. 4(e)], suggesting an energy gap at low temperatures. Figure 4(f) summarizes the leading-edge position as a function of the temperature, which shows the development of the gap below 130 K, consistent with the density-wave transition. The saturated gap at the lowest temperature $\Delta_0$ is about 12 meV, smaller than the previously reported value of 20 meV determined by the peak position in the flat band [35]. The reduced gap is estimated to be $\frac{2\Delta_0}{k_B T} \approx 2.1$, less than the mean-field expectation of 3.52, suggesting a relatively weak energy condensation of the density-wave state in $La_4Ni_3O_{10}$ [43].

The situation along the $\bar{\Gamma}\bar{X}$ direction is more complex due to the two nearly-degenerate bands crossing $E_F$ [Figs. 4(g-j)]. While the α band likewise shows a reduced spectral weight at low temperatures, the β band develops a peak below 130 K, as better visualized in the temperature-dependent EDCs in Figs. 4(k) and (l). Moreover, the leading edge of the EDCs at $k_F^\alpha$ along $\bar{\Gamma}\bar{X}$ shows a negligible shift compared to the $\bar{\Gamma}\bar{S}$ direction. This contrasting behavior along the two directions indicates an anisotropic CDW gap in the α band of $La_4Ni_3O_{10}$ at ambient pressure [39]. This behavior is likely related to the hybridization between the $d_{x^2-y^2}$ and $d_{z^2}$ orbitals, which is strongest along the $\bar{\Gamma}\bar{S}$ direction but becomes negligible along $\bar{\Gamma}\bar{X}$ [Fig. 3(c)]. At $k_F^\beta$, we observe a small leading-edge shift of about 6 ± 4 meV, compared to the peak position at about -20 meV at 20 K, which was not resolved in the previous measurement [35]. For the flat γ band, no shift with temperature is observed (Supplemental Materials [40]), which is in contrast to the previous report [35], but similar to the case in $La_3Ni_2O_7$ [17].

Our measurements of the ambient phase of the two compounds provide abundant information for understanding the novel normal-state properties and deciphering the unconventional superconductivity of RP-phase nickelates at high pressure. Firstly, the experimental low-energy electronic structures of the two compounds are very alike, which fit well to the calculated bands (Supplemental Materials [40]). This surprising observation suggests a similar electronic basis for the superconductivity in the two materials [44]. Secondly, the nice agreement between the calculated results and experiments at ambient pressure with an orbital-dependent renormalization can possibly be extrapolated to the pressurized phases, which is a necessary ingredient in theoretical modeling. In general, our data demonstrate the more correlated nature of the $d_{z^2}$ orbital than the $d_{x^2-y^2}$ orbital, consistent with the theoretical understanding of the distinctive correlation of the two orbitals due to electron filling [27,45,46]. Thirdly, with a flat band top near $E_F$, the FS sheets of the $d_{z^2}$ bands are highly sensitive to the exact position of $E_F$. In $La_3Ni_2O_7$, the band is proposed to be metalized with a new pocket in the FS under pressure [10], which is believed to be important to the HTSC [47,48]. Indeed, the subtle Fermiology can conspire with the oxygen deficiency to strongly influence the electron pairing, and even spoil the HTSC of $La_3Ni_2O_7$ [32,48-50]. Finally, the sharp density-wave transition shown in the ambient phase of $La_4Ni_3O_{10}$ might compete with the superconductivity, as the electronic DOS at $E_F$ is suppressed by the CDW transition. For $La_3Ni_2O_7$, by contrast, the phase transition seems to be weaker and more elusive [10,19,44,48], which might be an advantage for the HTSC.

In conclusion, we systematically investigate the electronic structure of $La_4Ni_3O_{10}$ and $La_3Ni_2O_7$ at ambient pressure both experimentally and theoretically, and explore the effect of pressure theoretically. The high similarity in the electronic structure of these two materials is revealed. The anisotropic energy gap related to the density-wave transition accords with the momentum distribution of the orbital components in the corresponding energy band in $La_4Ni_3O_{10}$. The nice agreement of the experimental results with the calculation after an orbital-dependent renormalization provides important hints to the exploration of HTSC in nickelates.

**Acknowledgement**

This work is funded by the National Key R&D Program of China (Grant No. 2022YFA1403201 and Grant No. 2022YFA1403100), the National Natural Science Foundation of China (No. 12274251, 12004270, and 52272265), and the Guangdong Basic and Applied Basic Research Foundation (Grant No. 2022B1515120020). L.X.Y. acknowledges support from the Tsinghua University Initiative Scientific Research Program and the Fund of Science and Technology on Surface Physics and Chemistry Laboratory (No. XKFZ202102). We thank the MAX IV Laboratory for the time on beamline Bloch under Proposal 20230668, and the SSRF for the time on beamline 03U under Proposal No. 2023-SSRF-PT-502880.

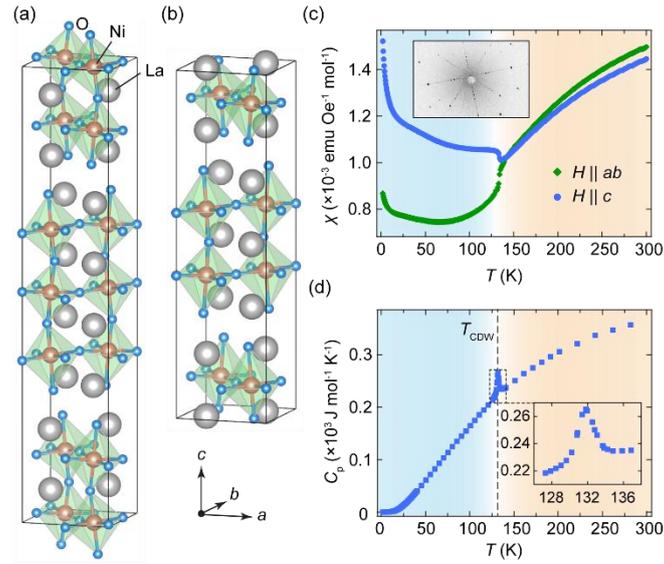

FIG. 1. (a, b) Crystal structures of La$_4$Ni$_3$O$_{10}$ (a) and La$_3$Ni$_2$O$_7$ (b) at ambient pressure. (c) Magnetic susceptibility along the *ab* and *c* directions as a function of temperature of La$_4$Ni$_3$O$_{10}$. The inset shows the Laue pattern for (001) surface. (d) Specific heat as a function of temperature of La$_4$Ni$_3$O$_{10}$. The inset highlights the cusp near 132 K.

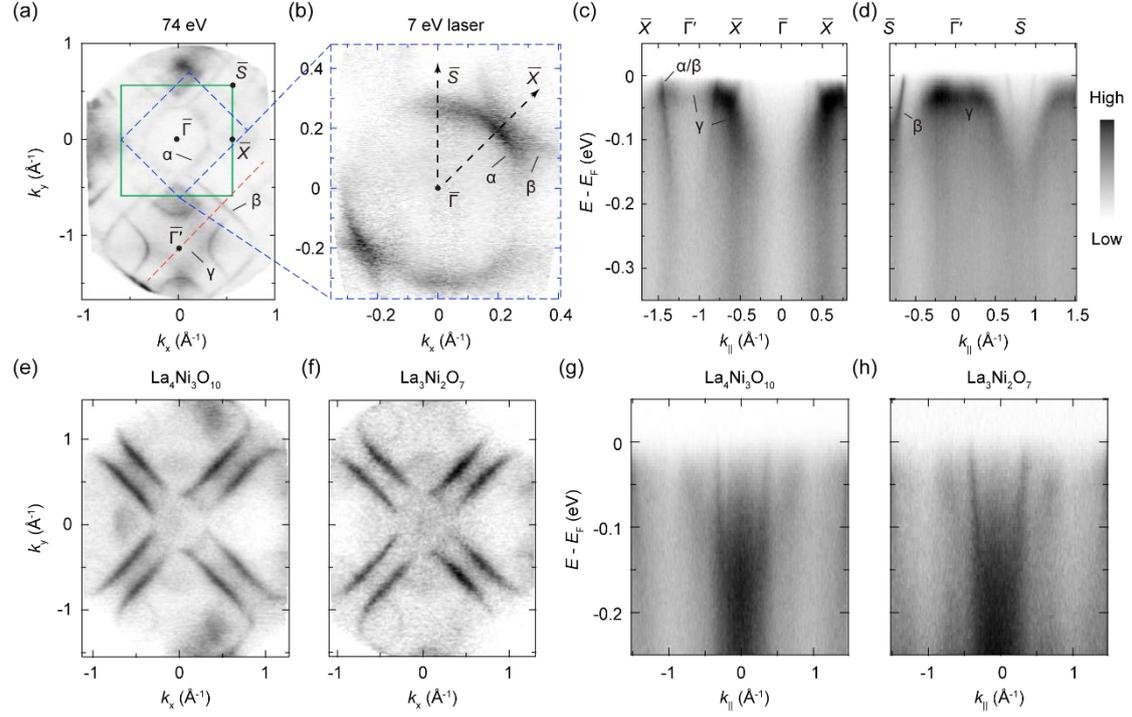

FIG. 2. (a) Experimental Fermi surface (FS) of $La_4Ni_3O_{10}$ integrated within an energy window of ± 20 meV around the Fermi level ($E_F$). Data were collected with 74 eV photons. The surface Brillouin zone (BZ) is overlaid as green lines. (b) FS measured with a 7-eV laser corresponding to the blue dashed area in (a). (c, d) Band dispersions along the high-symmetry directions of $\bar{\Gamma}\bar{X}$ (c) and $\bar{\Gamma}\bar{S}$ (d) measured at the photon energy of 74 eV. Data in (d) were collected with a geometry indicated by the red dashed line in (a). (e), (f) Experimental FS of $La_4Ni_3O_{10}$ and $La_3Ni_2O_7$ measured at the photon energy of 98 eV, respectively. (g, h) Band dispersions of $La_4Ni_3O_{10}$ and $La_3Ni_2O_7$ along $\bar{\Gamma}\bar{X}$ measured at the photon energy of 98 eV, respectively. Data in (b) were collected with linear-vertically polarized photons at 80 K. All other data were collected with linear-horizontally polarized photons at 20 K.

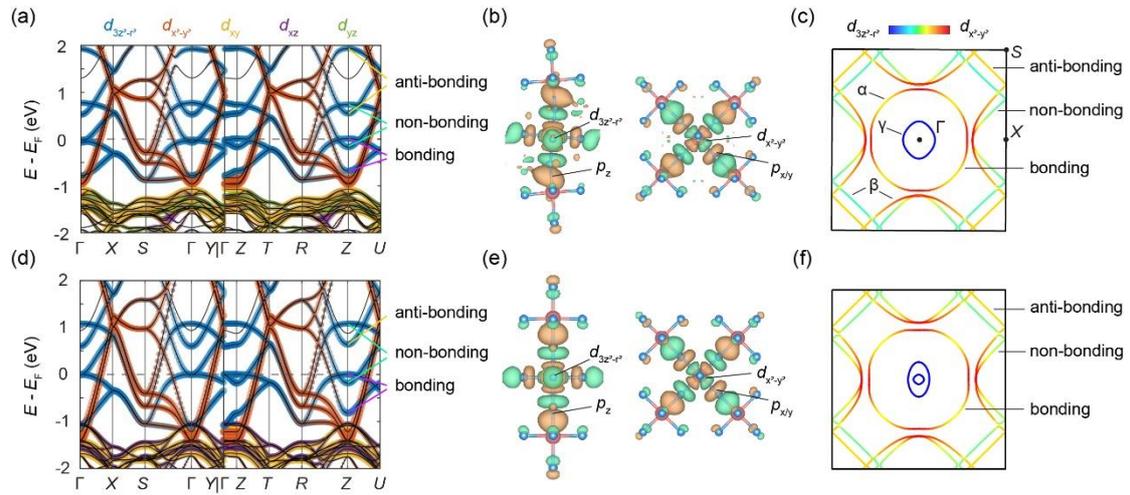

FIG. 3. (a) Calculated band structure with atomic orbital projection for the ambient phase of La$_4$Ni$_3$O$_{10}$. (b) Calculated Wannier orbital of the ambient phase showing the Ni $3d_{z^2}$–O $2p_z$ and Ni $3d_{x^2-y^2}$–O $2p_{x/y}$ hybridized atomic orbitals that are centered at Ni atoms in the middle Ni-O layers. (c) Cross-section of FS in the $k_z = 0$ plane with atomic orbital projections. (d-f) Same as those in (a-c), but for the high-pressure phase of La$_4$Ni$_3$O$_{10}$.

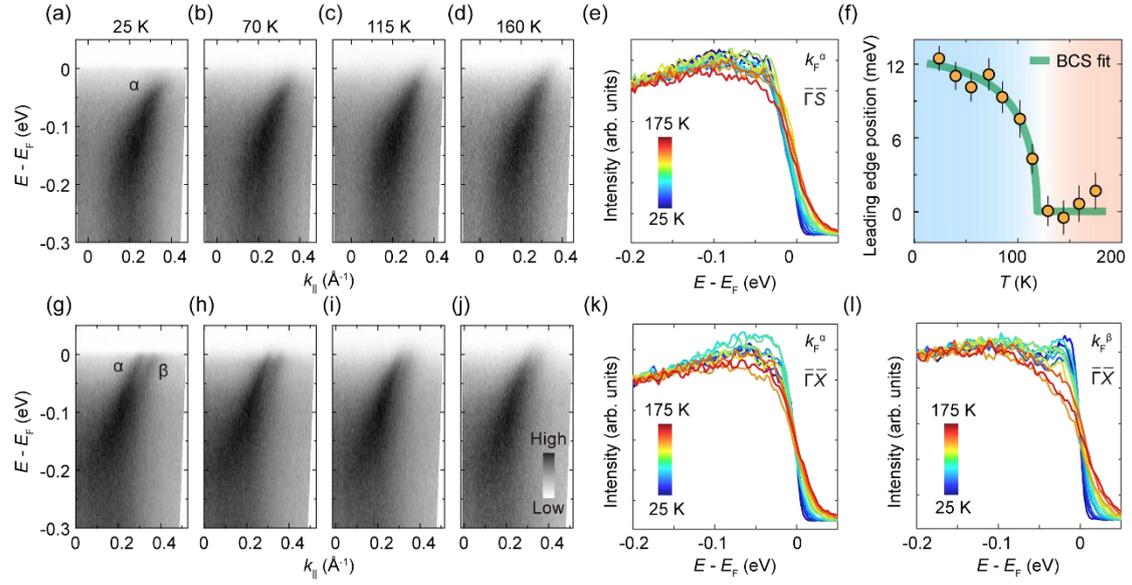

FIG. 4. (a-d) Temperature-dependent ARPES spectra of $La_4Ni_3O_{10}$ along the $\overline{\Gamma}\overline{S}$ direction. (e) The energy distribution curves (EDCs) at the Fermi momenta ($k_F$) at different temperatures. (f) Corresponding leading-edge positions of the EDCs. (g-j) Temperature-dependent ARPES spectra along the $\overline{\Gamma}\overline{X}$ direction. (k, l) The EDCs at $k_F^\alpha$ and $k_F^\beta$ at different temperatures, respectively. Data in (a-f) and (g-l) were collected with linear-vertically and linear-horizontally polarized photons with 7-eV energy, respectively.